\documentclass[aip,jcp,reprint]{revtex4-1}
\usepackage{amsmath}
\usepackage[utf8]{inputenc}
\usepackage{xcolor}
\usepackage{amsfonts}
\usepackage{todonotes}
\usepackage[version=3]{mhchem}

\usepackage[english]{babel}
\newcommand{\bigO}[1]{\ensuremath{ \mathcal{O}\left( #1 \right) }}

\usepackage[normalem]{ulem}
\newcommand\redout{\bgroup\markoverwith
{\textcolor{red}{\rule[.5ex]{2pt}{0.4pt}}}\ULon}
\usepackage{xcolor}

\newcommand{\mat}[1]{\ensuremath{\mathbf{#1}}}

\DeclareMathOperator{\rank}{rank}
\DeclareMathOperator{\area}{area}

\newcommand{\omat}[2]{\ensuremath{\mathbf{#1}^{\mathbf{#2}}}}

\newcommand{\trans}[1]{\ensuremath{\left( #1 \right)^{\dagger}}}

\newcommand{\lives}[3]{\ensuremath{\mathbb{#1}^{ #2 \times #3}}}

\newcommand{\chemints}[2]{\left( #1 | #2 \right)}

\newcommand{\epsil}[1]{\ensuremath{\epsilon_{\mathrm{#1}}}}

\newcommand{\CC}{C\nolinebreak\hspace{-.05em}\raisebox{.4ex}{\tiny\bf +}\nolinebreak\hspace{-.10em}\raisebox{.4ex}{\tiny\bf + }}

\begin{document}

\title{Clustered Low-Rank Tensor Format:
    Introduction and Application
    to Fast Construction of Hartree-Fock Exchange}
\author{Cannada~A.~Lewis}
\author{Justus~A.~Calvin}
\author{Edward~F.~Valeev}
\email{efv@vt.edu}
\affiliation{Department of Chemistry, Virginia Tech, Blacksburg,%
    Virginia 24061, USA}

\begin{abstract}
Clustered Low Rank (CLR) framework for block-sparse and block-low-rank tensor representation and computation is described. The CLR framework depends on 2 parameters that control precision: one controlling the CLR block rank truncation and another that controls screening of small contributions in arithmetic operations on CLR tensors. As these parameters approach zero CLR representation and arithmetic become exact. There are no other ad-hoc heuristics, such as domains. Use of the CLR format for the order-2 and order-3 tensors that appear in the context of density fitting (DF) evaluation of the Hartree-Fock (exact) exchange significantly reduced the storage and computational complexities below their standard \bigO{N^{3}} and \bigO{N^{4}} figures. Even for relatively small systems and realistic basis sets CLR-based DF HF becomes more efficient than the standard DF approach, and significantly more efficient than the conventional non-DF HF, while negligibly affecting molecular energies and properties.
\end{abstract}

\maketitle


\section{Introduction}

    Despite the exponential complexity of many-body quantum mechanics -- a
    manifestation of ``the curse of dimensionality'' -- many important classes
    of problems, such as electronic structure in chemistry and materials
    science, have robust polynomial solutions that become exact for practical
    purposes.\cite{Tajti:2004hwa,Bomble:2006dv,Harding:2008cn,
        Shiozaki:2009ix,Booth:2013er,Yang:2014jc} However, such solutions are
    limited by the high-order polynomial complexity of data and operations. For
    example, the straightforward implementation of CCSD\cite{Scuseria:1988dn} -- the
    coupled-cluster\cite{Bartlett:1981hg} model with 1-body and 2-body
    correlations -- has \bigO{N^6} and \bigO{N^4} operation and data
    complexities, respectively. This is significantly more expensive than the
    corresponding \bigO{N^4} and \bigO{N^2} complexities of hybrid Kohn-Sham
    density functional theory (KS DFT) that dominates chemistry applications.
    Thankfully it is possible to improve on these figures.

    Fast numerical algorithms trade off precision and/or small-$N$ cost for
    improved asymptotic scaling. A classic example is Strassen's algorithm for
    matrix multiplication\cite{Strassen:1969uw} that has higher operation
    count than the na\"ive algorithm for small matrices but is faster for large
    matrices, namely \bigO{N^{\log_{2} 7}} vs. \bigO{N^{3}}. Operation count of Strassen-based
    implementation of CCSD is therefore $\bigO{N^{2 \log_{2} 7}}
    \approx \bigO{N^{5.6}}$. Another notable fast algorithm of particular
    importance to the field of electronic structure is Fast Multipole Method
    (FMM)\cite{Greengard:1987gi, White:1994ip} 
    applies an integral (e.g.  Coulomb) operator in
    \bigO{N} operations, instead of 
    \bigO{N^{2}}, for any finite precision $\epsilon$.

    In molecular electronic structure fast algorithms are traditionally predicated
    on taking advantage of the sparse structure of the operators and
    states. Such structure in most cases takes form of the
    {\em element} or {\em block} sparsity. For example, the one-electron density matrix
    is conjectured\cite{Kohn:1996hn,Li:1993jra, Millam:1997cla}
    to decay exponentially in insulators when expressed in localized basis (AO
    or Wannier); this is the foundation of fast density matrix minimization
    in one-particle theories and also leads to the exponential decay of the ``exact'' (Hartree-Fock)
    exchange operator appearing in hybrid DFT and many-body methods.  Linear
    scaling methods based on such strategy have been demostrated by multiple
    groups.\cite{Ochsenfeld:2007vwa,Niklasson:2003hha,Shao:2003hya,Challacombe:1999bpa,Palser:1998uk,Millam:1997cla}
    While the element-sparsity-based strategy is appropriate for LCAO in small
    basis sets and low-dimensional systems, the sparsity of density matrix in
    three-dimensional systems is remarkably
    low,\cite{Rudberg:2008cba,Rudberg:2011km,VandeVondele:2012ho} especially
    when expressed in realistic (triple- and higher-$\zeta$) basis sets necessary
    for many-body methods and even hybrid DFT; e.g., the density matrix of a
    32000-molecule water cluster is only 83\% sparse (!)
    even when expressed in a double-$\zeta$ basis.\cite{VandeVondele:2012ho}
	Similar conclusions can be drawn from the early attempts to develop practical
	{\em many-body} methods solely by using sparsity of correlation operators
    in AO basis.\cite{Scuseria:1999bp} It is clear that the element sparsity
    alone is hardly sufficient
    for practical fast electronic structure.
	Another structure used in fast electronic structure methods
	is {\em rank} sparsity.
	For example, the Coulomb operator, $\hat{V} f(x) \equiv \int \mathrm{d}x \, f(x') / |x-x'|$,
	has a dense matrix representation due to slow decay of the integral kernel,
	but the rank of off-diagonal blocks when expressed in localized basis
	is low due to the smoothness of the kernel; of course, this is the basis of FMM.\cite{Greengard:1987gi}
	The key lesson here is that while globally the Coulomb operators has no non-trivial rank-sparsity,
	ranks of off-diagonal blocks are low. 
	Remarkably, similar local rank-sparsity is exploitable in other contexts. For example,
	blocks of many-body wave functions have low rank when expressed in localized basis;
	such rank-sparsity is the foundation of fast many-body methods
	based on local pair-natural orbitals (PNO)\cite{Neese:2009kpa}
	recently demonstrated in linear-scaling form.\cite{Riplinger:2013jk,Pinski:2015ii}

    In this work we propose a practical approach to recovering
    element and rank sparsities (termed here data sparsity)
    by using a general tensor format framework
    called Clustered Low-Rank (CLR, pronounced as ``CLeaR'').
    Some existing approaches, such as PNO-based
    many-body methods, can be viewed as a specialized application of
    CLR. However, we demonstrate in the following discussion that CLR has
    powerful applications beyond this particular context.
    
    The theoretical underpinnings of CLR come from the study of semi-separable
    \cite{Vandebril:2005fi,Chandrasekaran:2005kv}, hierarchically
    semi-separable\cite{Xia:2010kx}, and $\mathcal{H}$-
    matrices.\cite{Hackbusch:1999jh,Hackbusch:2000fx,Grasedyck:2003ij} Other
    researchers have mentioned this type of approximation, with regards to
    electronic structure, before, namely \onlinecite{Bock:2013ixa} where the
    idea was only mentioned in passing and  very recently
    \onlinecite{benner2015reduced} where a rank structured approach was used to
    reduce computational complexity in the solution to the Bethe-Salpeter
    equation.
    These can be viewed as algebraic generalizations of the rank-sparse
    structure arising in integral boundary problems, studied first by
    Rokhlin\cite{Rokhlin:1985ig} and Hackbusch\cite{Hackbusch:1989tn}. Later
    Tyrtyshnikov realized that this type of approximation could be generalized
    to matrices arising in integral
    equations via a method he calls the mosaic-skeleton approximation (MSA)
    where the integral kernel satisfies some necessary
    conditions.\cite{Tyrtyshnikov:1996fd} He expands on this idea suggesting
    that for matrices that fit the necessary conditions there exists a
    non-overlapping blocking called a mosaic in which many of the blocks have
    reduced rank.\cite{tyrtyshnikov1997bookchap,Goreinov:1997hh} Soon after
    Hackbusch premiered the $\mathcal{H}$-matrix concept, where $\mathcal{H}$
    stands for hierarchical, in which he formalizes his integral approximation
    methods into a general matrix
    framework. The main difference
    between Tyrtyshnikov and Hackbusch's generalizations are that
    $\mathcal{H}$-matrices have a hierarchical block structure while the MSA
    approach does not. In this regard CLR can be viewed as a refinement of
    MSA for general tensor data; the lack of hierarchy is deemed important for compatibility
    with standard algorithms for matrix and tensor algebra in high-performance
    computing that embed blocked tensors onto a regular cartesian grid of
    processors.

    The rest of the paper is organized as follows: in section \ref{sec:CLR} we
    will discuss the details of CLR, in section \ref{sec:clr_app} we discuss an
    application of CLR to density fitting approximation in the context of the Hartree-Fock (exact) exchange
    evaluation, and in section \ref{sec:conclusions} we
    discuss our findings as well as future directions for CLR.
    
\section{\label{sec:CLR} Clustered Low Rank Approach}
    CLR, in a general sense, may be thought of as a tensor representation 
    {\em framework}, which divides a tensor into sub-blocks (tiles) that are
    stored in either low-rank or full-rank (dense) form.  For example, consider
    a real-valued, order-$k$ tensor with dimension sizes 
    $\{ n_{1} \dots n_{k} \}$:
    \begin{align}
         {\bf T} : \mathbf{x} \to \mathbb{R}, \quad 
             \forall \mathbf{x} \equiv \{x_{1} \dots x_{k} \}, x_{i} 
             \in [1,n_{i}] \, \forall i=1,k ,
    \end{align}
    with tensor elements denoted as 
    $T_{\mathbf{x}} \equiv T_{x_{1} \dots x_{k}} = {\bf T}(\mathbf{x})$.
    CLR representation of {\bf T} is defined as
    follows:
    \begin{enumerate}
        \item Each dimension $i$ is tiled into $\nu_{i} \geq 2$ blocks (tiles) of sizes $\{ \beta_{1} \dots \beta_{\nu_{i}} \}$:
        \begin{align}
          \tau_{i} : & \{ 1 \dots n_{i} \} \to \{ \mathcal{X}^{(i)}_{\xi}\}, \xi\in[1,\nu_{i}], \quad \mathrm{where} \\
          \mathcal{X}^{(i)}_{\xi} & \equiv \{ x^{(\xi)}_{1} \dots x^{(\xi)}_{\beta_{j}} \}
        \end{align}
        Tiling of a tensor domain is then obtained as a tensor product of dimension tilings,
        $\tau \equiv \tau_{1} \otimes \dots \tau_{k}$ so that
        each block index $\boldsymbol{\xi} \equiv \{ \xi_{1} \dots \xi_{k} \}$
        defines block of element indices $\mathcal{X}_{\boldsymbol{\xi}} \equiv \mathcal{X}^{(1)}_{\xi_{1}} \otimes \dots \mathcal{X}^{(k)}_{\xi_{k}}$. The corresponding tensor tile,
        \begin{align}
        {\bf T}_{\boldsymbol{\xi}} \equiv {\bf T}_{\mathcal{X}^{(1)}_{\xi_{1}} \dots \mathcal{X}^{(k)}_{\xi_{k}}} = \{ T_{x_{1} \dots x_{k}}: x_{j} \in \mathcal{X}^{(j)}_{\xi_{j}}, \forall j\in[1,k] \},
        \end{align}
        is also an order-$k$ tensor.
        \item The (optional) {\em shape} predicate $z(\boldsymbol{\xi})$ defines whether block $\boldsymbol{\xi}$ is set to zero.
        \item Each tensor block ${\bf T}_{\boldsymbol{\xi}}$ is data-compressed using a low-rank tensor decomposition.
    \end{enumerate}
    
    It is evident from the above description that CLR is similar to 
    $\mathcal{H}$-matrices and MSA with respect to the structure
    of the matrix representations. However, the choice of tiling,
    shape, and low-rank decomposition schemes as well as the 
    specific application chemical knowledge differentiates these
    methods. In the following sections, we will specialize CLR for a
    particular set of matrices. But first we should discuss the 
    design elements of CLR framework.
    \begin{itemize}
    \item Basis elements are reordered into spatially localized 
        clusters that form the non-overlapping tile partitions of CLR
        matrices (see Section~\ref{sec:k-means} for details). This
        tiling structure is essential for both improved data locality in
        the context of parallel computing,\cite{Weber:2015kr} and
        data compression in low-rank tiles.
    \item The {\em shape} defines block-sparse structure of CLR
        matrices, which may be based on the magnitude of tile norms
        (or norm estimates), an {\em imposed} sparsity model, or a
        combination of these methods. Many useful electronic
        structure methods impose block sparsity, {\em e.g.}\ the use
        of AO domains in local density fitting and local correlation.
        Although low-rank decomposition of blocks naturally reveals 
        block sparsity, the preemptive introduction of block sparsity
        can lead to additional computational savings. 
    \item Low-rank decomposition of tensor tiles is used to reduce
        the total storage requirements for high-order (greater than 
        order-two) tensors. The choice of decomposition methods
        is problem specific due to the fact that, with the exception
        of matrices, tensor rank is not uniquely defined, nor is there
        an optimal decomposition for tensors.  Even for matrices,
        SVD is often not the optimal choice due to its high 
        computational cost.
    \item Another factor in the choice of decomposition is the
        suitability an algorithm for use in algebraic operations,
        where frequent recompressions will often be necessary. For
        example, given a matrix, SVD will always provide the most
        compact low-rank representation for a given accuracy, but it
        may be beneficial to sacrifice storage or accuracy in certain
        cases for computational efficiency.
    \end{itemize}

    To demonstrate the usefulness of the CLR framework in practical 
    applications, we applied it to fast construction of the Hartree-Fock (exact) exchange
    using density fitting (DF) as our first example. DF (also known as resolution of the identity) technology
    allows us to reduce the computational cost of the Fock operator construction
    by decreasing the prefactor, especially when triple- or
    higher-$\zeta$ basis sets are used. However, for large molecules DF HF
    performs poorly, relative to conventional HF methods, due to 
    the \bigO{N^{3}} storage and \bigO{N^{4}} operation complexities. Meanwhile
    even the na\"ive implementation of conventional HF methods only require
    quadratic storage, and \bigO{N} approaches to exchange build in LCAO representation
    are well-known.\cite{Schwegler:1996hh,Burant:1996cf,Ochsenfeld:1998p1663,Neese:2009ca} The goal of our work is
    therefore to develop a CLR-based DF HF method with reduced
    storage and computation complexities.
    
    The DF HF method deals with at most order-3 tensors, e.g. the so-called three-center two-electron Coulomb integral:
    \begin{align}
    \label{eq:3c2e-int}
     (\kappa|\mu \nu) &\equiv \iint \phi_{\kappa}(\mat{r}_1)  \frac{1}{r_{12}}
            \phi^{*}_\mu(\mat{r}_2) \phi_\nu(\mat{r}_2)   \, d\mat{r}_1 d\mat{r}_2 \text{.}
    \end{align}
    An effective low-rank representation of such tensors can be obtained by
    treating its blocks as matrices, where $\mu$ and $\nu$ forms a
    single index space.  Hence, we describe in detail use of CLR for compressed
    representation and operation of matrices and matrix-like tensors.

\subsection{Basis Clustering} \label{sec:k-means} 
 
    To tile AOs as well as occupied molecular orbitals (MO)
    in this work we utilized the k-means clustering
    algorithm\cite{Lloyd:1982do, Jain:2010cu} with a Euclidean distance metric.
    Given a set of Cartesian coordinates $\{\mat{r}_1, \mat{r}_2, \dots,
    \mat{r}_m\}$ in $\mathbb{R}^{n}$ (atomic coordinates
    when clustering atoms/AOs, or the expectation values of $\hat{\mat{r}}$ when clustering MOs)
    and the target number of clusters $m$, k-means seeks clusters that minimize
    \footnote{An incorrect objective function of 
        $\sum_{i = 1}^m \sum_{\mat{r} \in \mathcal{X}_{i}} || \mat{r} -
        \mat{R}_{i} ||$ was used to evaluate which k-means++ clustering was
        best. Thus it is possible that the chosen clustering was not the
        minimal one from the initial guesses, but it is still a local minimum.
        Only the clustering in the auxiliary basis was affected since other
        bases used natural blocking. We expect this oversight will be of no
        practical consequences.
    }
    \begin{align}
    \label{eq:kmeansobjectivefunction}
        \sum_{i = 1}^m \sum_{\mat{r} \in \mathcal{X}_{i}} 
            || \mat{r} - \mat{R}_{i} ||^2 \text{,}
    \end{align}
    where $\mat{R}_i$ is the center of mass of cluster $\mathcal{X}_{i}$ for
    clustering atoms or the centroid when clustering MOs.
    To determine the clusters, we use
    the k-means++ algorithm\cite{arthur2007k} to seed starting points, and
    then proceed to perform the standard k-means; the only exception is that
    hydrogen atoms are always forced into the same clusters as their nearest non-hydrogen atom.
    We stop when all centers of mass reach a (local) minima, or after 100
    iterations.  The procedure is repeated 10 times with random starting seeds.
    From these 10 clusterings we use the one that has the lowest value
    for the objective function, \eqref{eq:kmeansobjectivefunction}.

\subsection{Block Sparsity in CLR Tensors}

    Tensors like that in equation \eqref{eq:3c2e-int} become increasingly
    sparse for large molecules when expressed in a localized basis.
    Thus, it is mandatory to estimate norms of the blocks
    to avoid evaluation of vanishing blocks. For the Coulomb 3-index tensor \eqref{eq:3c2e-int}
    in an AO basis, this is readily done
    by using AO shell-wise integral estimators, {\em e.g.}\ 
    ref \onlinecite{Hollman:2015ca}.
    An extension to atom and multi-atom blocks is straightforward.
    The block-norm estimates define the sparsity of the tensors (see 
    section~\ref{sec:blk_sp_arithmatic}  for details).
    For simplicity and clarity in this work, we did not preemptively skip computation of any integrals.
    This will be addressed in future work.

\subsection{Low-Rank Block Representation and Arithmetic}
    Each block of a CLR matrix is potentially represented in a low-rank form.
    Before discussing multiplication and addition of CLR matrices, we must
    first consider the block-level operations in low-rank representation.
    
\subsubsection{Low-Rank Matrix Approximation}

    Matrix $\mat{A}_{r} \in \lives{R}{n}{m}$ has rank $r \leq \min{(n,m)}$ if
    it can be represented {\em exactly} as a sum of outer products of no fewer
    than $r$ linearly independent vectors 
    \begin{align} \label{eq:lr_vec_decomp}
        \mat{A}_{r} &= \sum_{i=1}^{r} \mat{s}_{i}^{\mat{A}} 
            \trans{\mat{t}_{i}^{\mat{A}}}\text{,} \\
	       & \equiv \label{eq:lr_decomp}
        \mat{S}^{\mat{A}} \trans{\mat{T}^\mat{A}}\text{,}
    \end{align}
    where $\mat{S}^{\mat{A}} \equiv \{ \mat{s}_{1}^{\mat{A}},
    \mat{s}_{2}^{\mat{A}}, \dots \mat{s}_{r}^{\mat{A}} \}$ and $\mat{T}^{\mat{A}}
    \equiv \{ \mat{t}_{1}^{\mat{A}}, \mat{t}_{2}^{\mat{A}}, \dots
    \mat{t}_{r}^{\mat{A}} \}$.  We seek to approximate a
    full-rank matrix $\mat{A}$ with a rank-$r$ matrix $\mat{A}_{r}$ which both 
    minimizes 
    \begin{align}
        || \mat{A} - \mat{A}_r ||_{F} \leq \epsil{lr}
    \end{align}
    and allows for efficient arithmetic in decomposed form.
    Therefore, we approximate matrices using column pivoted (rank-revealing)
    QR decomposition (RRQR). RRQR is
    faster than SVD and approximates the SVD (optimal) ranks
    well.\cite{Chan:1987tq,QuintanaOrti:2006df,Bischof:1998hc} 
    The RRQR decomposition decomposes a matrix $\mat{A}$ as:
    \begin{align}
        \mat{AP} &= \mat{QR} \equiv \mat{Q}\begin{pmatrix}R_{11} & 
            R_{12} \\ 0 & R_{22} \end{pmatrix}\text{.}
    \end{align}
    The singular values of the matrix may be estimated, by computing
    $||\mat{R}_{22}||_F$, according to: 
    \begin{align} \sigma_{m - r+1} \leq
        ||\mat{R}^r_{22}||_2 \leq ||\mat{R}^r_{22}||_F \text{,}
    \end{align} 
    for $\mat{R} \in \lives{R}{m}{m}$. To estimate the rank of $\mat{A}$ we
    compute $\mat{R}$ (using the DGEQP3 LAPACK function) and walk up the lower
    triangle from element $R_{mm}$ to compute norms $||\mat{R}^r_{22}||_F$,
    stopping once the norm is larger than our threshold. While this method will
    potentially lead to larger ranks that using the SVD decomposition, in
    practice RRQR does a adequate job of revealing rank while having a reduced
    prefactor.
    
    Another consideration is that we consider a matrix low rank if for
    threshold \epsil{lr} matrix $\mat{A} \in \lives{R}{m}{n}$ has rank $r_A <
    \text{min}\left( m, n\right)$, but just because a matrix is low rank
    does not mean that there are computational or storage savings. Given a
    square matrix in $\lives{R}{m}{m}$ it is necessary that $r <
    \frac{m}{2}$ for there to be storage savings from storing $\mat{S}$ and
    $\mat{T}$ for the matrix.  Thus any time the storage of $\mat{S} + \mat{T}$
    is greater than the storage of $\mat{C}$ we convert the tensor to full rank
    form.

\subsubsection{Multiplication of Low-Rank Matrices}
    Clearly, $\mat{C} = \mat{A} \mat{B}$
    can be directly computed in low-rank form
    $\mat{C} = \mat{S}^\mat{C} \trans{\mat{T}^\mat{C}}$ using low-rank representations of \mat{A} and \mat{B}
    with rank
    $\min(\rank(\mat{A}),\rank(\mat{B}))$:
    \begin{align}
        \mat{S}^{\mat{C}} \trans{\mat{T}^{\mat{C}}} = &
            \mat{S}^{\mat{A}} \trans{\mat{T}^{\mat{A}}} 
            \mat{S}^{\mat{B}} \trans{\mat{T}^{\mat{B}}} \\
            = & \begin{cases}
            \mat{S}^{\mat{A}} \left( \left( \trans{\mat{T}^{\mat{A}}} 
            \mat{S}^{\mat{B}}\right)  \trans{\mat{T}^{\mat{B}}} \right)  & r_A < r_B \\
            \left( \mat{S}^{\mat{A}} \left( \trans{\mat{T}^{\mat{A}}} 
            \mat{S}^{\mat{B}}\right) \right)  \trans{\mat{T}^{\mat{B}}} & r_A \geq r_B 
        \end{cases}
        ,
    \end{align} 
    in which the order of evaluation and hence the definitions of
    $\mat{S}^{\mat{C}}$ and $\mat{T}^{\mat{C}}$ are specified by the
    parentheses. Note that the multiplication never increases matrix rank.

    The special cases where either $\mat{A}$ or $\mat{B}$ is full rank are also
    completely straightforward.

\subsubsection{Addition of Low-Rank Matrices}
	
	Unlike the multiplication, addition of low-rank matrices may increase rank
	as can be immediately seen:
    \begin{align} \label{eq:lf_sum}
        \mat{C} = & \mat{S}^{\mat{A}} \trans{\mat{T}^{\mat{A}}} 
            + \mat{S}^{\mat{B}} \trans{\mat{T}^{\mat{B}}} \nonumber \\
            = & \mat{S}^{\mat{A+B}} \trans{\mat{T}^{\mat{A+B}}} 
    \end{align}
    where $\mat{S}^{\mat{A+B}} = \{ \mat{S}^{\mat{A}} , \mat{S}^{\mat{B}} \}$
    and $\mat{T}^{\mat{A+B}} = \{ \mat{T}^{\mat{A}} , \mat{T}^{\mat{B}} \}$.
    The maximum rank of \mat{C}
    in \eqref{eq:lf_sum} is the sum of the ranks of \mat{A} and \mat{B}
    if columns of $\mat{S}^{\mat{A+B}}$ and/or $\mat{T}^{\mat{A+B}}$ are linearly independent.
    Of course it is often the case that the ranks of these matrices can be reduced further
	with controlled error. We can reduce the ranks of
    $\mat{S}^{\mat{A+B}}$ and $\mat{T}^{\mat{A+B}}$ as follows.
    Starting with QR decompositions of these matrices,
    \begin{align}
        \mat{S}^{\mat{A+B}} &= \omat{Q}{S} \omat{R}{S} \\
        \mat{T}^{\mat{A+B}} &= \omat{Q}{T} \omat{R}{T} \text{,}
    \end{align}
	\eqref{eq:lf_sum} is rewritten
    \begin{align}
        \mat{C} &= \omat{Q}{S} \omat{R}{S} \trans{\omat{R}{T}}
            \trans{\omat{Q}{T}} \text{,}
    \end{align}
    Intermediate $\mat{M} \equiv \omat{R}{S} \trans{\omat{R}{T}}$
    is then RRQR decomposed resulting in reduced rank $r \leq \rank\mat{A} +
    \rank\mat{B}$.
   \begin{align}
       \mat{M} = \widetilde{\mat{Q}} \widetilde{\mat{R}} 
           \overset{\epsil{lr}}\approx \widetilde{\mat{Q}}_r 
           \widetilde{\mat{R}}_r 
   \end{align}
   where for $\widetilde{\mat{Q}}$ and $\widetilde{\mat{R}}$ the tilde
   signifies RRQR as oppose to traditional QR.
   
   The final result for the low-rank form of $\mat{C}$ is
   \begin{align} \label{eq:lr_rounded_add}
       \mat{C} &= \left( \omat{Q}{S} \widetilde{\mat{Q}}_r \right) 
           \left(\widetilde{\mat{R}}_r \trans{\omat{Q}{T}} \right) 
           \equiv \mat{S}^C \trans{\mat{T}^C} \text{,}
   \end{align}
   where
   \begin{align}
       \mat{S}^C &= \omat{Q}{S} \widetilde{\mat{Q}}_r \\
       \trans{\mat{T}^C} &= \widetilde{\mat{R}}_r \trans{\omat{Q}{T}}.
   \end{align}
    if after compression $\rank\mat{C} \geq \frac{\rm full rank}{2}$ then
    $\mat{C}$ is converted to its full rank representation since no space
    savings is available.

   When performing addition if exactly one of  $\mat{A}$ or $\mat{B}$ is in the
   full rank representation, we perform the addition using a generalized
   matrix-matrix multiplication (GEMM). For example if $\mat{A}$ is low rank
   and $\mat{B}$ is full rank the addition proceed as follows:
   \begin{align} \label{eq:gemm_add}
       \mat{C} = \mat{S}^{\mat{A}} \trans{\mat{T}^{\mat{A}}} + \mat{B} \text{,}
   \end{align}
   where $\mat{C}$ is in its full rank representation. The addition in equation
   \eqref{eq:gemm_add} is able to take advantage of optimized linear algebra
   libraries and avoid the storage overhead of constructing a temporary of
   $\mat{A}$.

\subsection{Block-Sparse Arithmetic with CLR Tensors} \label{sec:blk_sp_arithmatic}

    An arithmetic operation on matrices/tensors composed of CLR-format
    blocks can be performed as a standard operations on dense matrices/tensors
    with standard block operations replaced by their CLR specializations.
    For the sake of computational efficiency, however, it is necessary to screen
	arithmetic expressions involving CLR-format blocks to avoid laboriously
	computing blocks that have small norm and hence low rank.
    Therefore we represent CLR matrices/tensors
    as their block-sparse counterparts, in which only some blocks
    are deemed to have non-zero norms; the surviving blocks are represented using
    CLR format. This allows us to exploit both block-sparsity and block-rank-sparsities.
	In the limit $\epsil{lr} \to 0$, CLR matrices/tensors become their ordinary block-sparse
	counterparts. Here we describe how we utilize block-sparse structure of CLR tensors
	in arithmetic operations.

    Block ${\bf A}_{\boldsymbol{\xi}}$ of CLR tensor ${\bf A}$ that is the result of an arithmetic operation
    is {\em nonzero} (hence, it will be computed) if its Frobenius norm satisfies
    \begin{align}
    \label{eq:eps-sp-def}
        \| {\bf A}_{\boldsymbol{\xi}} \|_{\rm F} \geq \epsilon_{\rm sp} \area{{\bf A}_{\boldsymbol{\xi}}}
    \end{align}
    where $\epsilon_{\rm sp}$ is a non-negative threshold, 
    and $\area {\bf A}_{\boldsymbol{\xi}}$ is the block {\em area}, i.e.
    the number of elements in the block.
    Because the area of each block within a tensor may vary significantly,
    the sparsity threshold is scaled by the block area to provide a consistent
    sparsity criteria with a single parameter.
    
    The sub-multiplicative property of Frobenius norm is utilized to estimate the norm
    in addition and contraction operations, e.g. for addition
    \begin{align} \label{eq:norm_add}
        \| {\bf A}_{\boldsymbol{\xi}} + {\bf B}_{\boldsymbol{\xi}} \|_{\rm F} \le 
            \| {\bf A}_{\boldsymbol{\xi}} \|_{\rm F} + \| {\bf B}_{\boldsymbol{\xi}} \|_{\rm F}
    \end{align}
    the upper bound provided by the RHS is used to estimate whether computing the block is warranted
    according to \eqref{eq:eps-sp-def}.
    Similarly, the norm of a contraction of two tensor blocks is bounded as
    \begin{equation} \label{eq:norm_cont_pair}
        \begin{split}
        \| {\bf A}_{\xi_i \dots \xi_k \dots} & {\bf B}_{\xi_k \dots \xi_j \dots} \|_{\rm F} \\
            \le & \| {\bf A}_{\xi_i \dots \xi_l \dots} \|_{\rm F} \| {\bf B}_{\xi_l \dots \xi_j \dots} \|_{\rm F}
        \end{split},
    \end{equation}
    where $\xi_i \dots$ and $\xi_j \dots $ are the outer block indices of the
    left- and right-hand tensors, respectively, and $\xi_k \dots$ is the
    set of block summation indices.
    Eqs. \eqref{eq:norm_add}, and \eqref{eq:norm_cont_pair} are combined to
    estimate the norm of the result blocks in a contraction operations:
    \begin{equation} \label{eq:norm_cont}
        \begin{split}
        \| {\bf C}_{\xi_i \dots \xi_j \dots} \|
            \le \sum_k \| {\bf A}_{\xi_i \dots \xi_k \dots} \|_{\rm F} \| {\bf B}_{\xi_k \dots \xi_j \dots} \|_{\rm F}
        \end{split}
    \end{equation}
    where ${\bf C}_{i \dots j \dots}$ is equal to $\sum_{k \dots} 
    {\bf A}_{i \dots k \dots} {\bf B}_{k \dots j \dots}$.
    Note that we do not screen {\em individual} contributions to the result blocks,
    i.e. all contributions are computed
    if the estimated result block norm satisfies Eq. \eqref{eq:eps-sp-def}.
    
    Additionally,
    we avoid explicit computation of norms of low-rank blocks by using the 
    multiplicative property of Frobenius norm:
    \begin{align}
         \| {\bf A}_{\boldsymbol{\xi}} \| 
             \approx \|{\bf S}^{\bf A}_{\boldsymbol{\xi}} ({\bf T}^{\bf A}_{\boldsymbol{\xi}})^{\dagger}\|_{\rm F}
             \le \|{\bf S}^{\bf A}_{\boldsymbol{\xi}}\|_{\rm F} \|{\bf T}^{\bf A}_{\boldsymbol{\xi}}\|_{\rm F} \text{.}
    \end{align}
    
    It is clear that computing norm estimates in complex expressions using upper bounds
    can quickly lead to poor norm estimates. Therefore, it is sometimes necessary to recompute the block norms and/or truncate
    {\em zero} blocks that are the result of computation.
    Similarly, after a CLR tensor contraction or addition operation it is
    possible that the rank of certain blocks are larger than necessary.
    Therefore in this work we sometimes recompute block norms and recompression
    after tensor contractions to minimize the ranks and reduce the
    storage.

\section{\label{sec:clr_app} Application: CLR-based Density Fitting in Hartree-Fock Method}

\subsection{Density Fitting and Hartree-Fock Method}
	Density fitting (also known as the resolution-of-the-identity)\cite{Whitten:1973ju,Baerends:1973p41,Feyereisen:1993di,Vahtras:1993wy,Dunlap:2000ii}
	in the context of electronic structure involves
	fitting the electron density or more often any product of
	(one-electron) functions as a linear combination of basis functions
	from a fixed (auxiliary) basis set to minimize some functional.
	This allows us to compute standard 4-center 2-electron
	integrals in terms of 2- and 3-center integrals:
    \begin{align}
        \chemints{\mu \nu}{\rho \sigma} &\approx \sum_{QP} E_{Q, \mu \nu}
            \left(\mat{V}^{-1}\right)_{Q,P} E_{P, \rho \sigma} \\
        & = \sum_{X} B_{X, \mu \nu} 
            B_{X,\rho \sigma} \text{,} \label{eq:b_expression} \\
        B_{X, \mu \nu} &\equiv \sum_{Q} E_{Q, \mu \nu} 
            \left(\mat{V}^{-1/2}\right)_{Q,X}
    \end{align}
	where $E_{Q,\mu\nu} \equiv
    \chemints{Q}{\mu \nu}$ (defined in Eq. \eqref{eq:3c2e-int})
    and $V_{P,Q} \equiv (P|Q)$ are the three- and two-center
    Coulomb integrals. In practice, instead of $\mat{V}^{-1/2}$ we use the inverse of
    Cholesky decomposition of \mat{V}; this is cheaper than computing the square root inverse
    and incidentally leads to better CLR compression in $\mat{B}$. 
	
    There are many different strategies for using density fitting in
    Hartree-Fock. Here we pick a basic approach that instead of the density matrix uses occupied
    MOs (or any other factorization of the density)\cite{Kendall:1997kh}
    (additional improvements such as the use of local
    density fitting\cite{Gallant:1996ua, Snijders:1998jr,
        Watson:2003eo,Merlot:2013ep,Hollman:2014gc}
    and computing only the occupied
    section of exchange matrix\cite{Manzer:2015de}
    can be also incorporated in our CLR-based scheme):
    \begin{enumerate}
        \item Compute and store $\mat{B}$
        from equation \ref{eq:b_expression}. 
        \item Form $\mat{J}$, the coulomb 
            contribution to the Fock matrix in \bigO{n^3} steps via 
            \begin{align}
                J_{\mu \nu} &= \sum_{X} B_{X, \mu \nu} 
                    \sum_{\rho \sigma} B_{X, \rho \sigma} D_{\rho \sigma}
            \end{align}
            where $\mat{D}$ is the AO density matrix. 
        \item Compute an intermediate 
            tensor $\mat{W}$ with one index transformed from the AO to MO space as 
            follows 
            \begin{align}
                W_{X, \mu i} &= \sum_{\sigma} B_{X, \mu \sigma} C_{\sigma i}
            \end{align}
            where the column vectors of $\mat{C}$ are the occupied molecular
            orbitals. 
        \item Form the exchange contribution to the Fock matrix
            \begin{align}
                K_{\mu \nu} &= \sum_{Xi} W_{X, \mu i} W_{X, \nu i} \text{.}
            \end{align}
    \end{enumerate}
    The two most expensive steps in forming the Fock matrix this way are 
    formation of $\mat{W}$ and $\mat{K}$, which both require approximately 
    $\mathcal{N} n^2 o$ steps where $\mathcal{N}$ is the size of the auxiliary 
    basis and $o$ is the number of occupied orbitals in the system.

    Although the use of density fitting in Hartree-Fock method has the same
    formal complexity as conventional Fock operator construction, namely $\bigO{N^{4}}$ due to the exchange computation,
    there is significant reduction in prefactor for large orbital basis (triple-$\zeta$ and larger)
    due to avoiding the relatively expensive and repeated (once-per-iteration) computation of
	4-center integrals.
	These advantages are exacerbated by the poor suitability of highly-irregular
	integral kernels to modern wide-SIMD architectures; in contrast, the dense matrix arithmetic
	of Eq. \eqref{eq:b_expression} can perform close to peak on most modern architectures provided the
	dimension sizes are large.
	
    Despite the cost advantages of DF, its cubic storage requirements and
    quartic cost prohibit direct application to large systems.  {\em Local}
    density fitting reduces the storage and computational complexity of density
    fitting by expanding localized products in terms of auxiliary basis
    functions that are ``nearby'' in some sense (geometric or
    otherwise)
    Local DF approximations can be difficult to make robust, thus in this work
    we decided to investigate whether a black-box reduced-scaling density
    fitting methodology can be obtained by the use of CLR format in the context
    of the Hartree-Fock method.

\subsection{CLR-based Density Fitting Hartree-Fock}

    Implementation of CLR-based DF HF method used the 4-step formulation listed
    above.  All dimensions were blocked naturally except for the index
    corresponding to the auxiliary basis, which uses larger blocks for
    performance reasons. Natural blocking consists of one block per water for
    water clusters and one block per CH$_2$ or CH$_3$ for n-alkanes.  For the
    auxiliary basis we set the number of clusters to be half the number of
    natural clusters and use our k-means algorithm described previously in
    \ref{sec:k-means} to determine clusters.
    Blocks of all three-index tensors were represented as low-rank matrices by
    merging the orbital indices together; the details of tensor arithmetic in
    this representation were described in Section \ref{sec:CLR}.
    
	To simplify the initial implementation we use a cubically-scaling eigensolve-based solver for the Hartree-Fock;
	we will switch to a \bigO{N} density minimization in the future.
	Instead of using occupied eigenorbitals directly we utilized orbitals obtained by pivoted
	Cholesky decomposition of the density matrix (LAPACK function DPSTRF), which are known to be more localized,\cite{Aquilante:2006dq}
    to increase the data sparsity of tensor $\mat{W}$.

    The CLR DF HF method was implemented with the help of the open-source {\sc TiledArray} parallel tensor
    framework.\cite{TiledArray}

\subsection{\label{sec:results} Results}

    To gauge the robustness of CLR DF HF we computed energies and electric dipole moments of
    quasi-1-d (n-alkanes with linear geometry) and 3-d (water clusters) systems. Cartesian geometries for
    n-alkanes were generated using Open Babel.\cite{OLBoyle:2011wz}
    Cartesian coordinates for the water clusters were taken from ErgoSCF's public repository\cite{ErgoWaterClusters};
    these clusters are random snapshots of molecular dynamics
    trajectories and are representative of liquid water structure
    at ambient conditions.\cite{Rudberg:2008cba}
    
    Dunning's correlation consistent basis sets and the matching auxiliary
    basis sets were utilized throughout this work.\cite{Dunning:1989bx,Weigend:2002jp}

\subsubsection{CLR Error Assessment}
    \begin{figure}[h!]
        \includegraphics[width=\columnwidth]{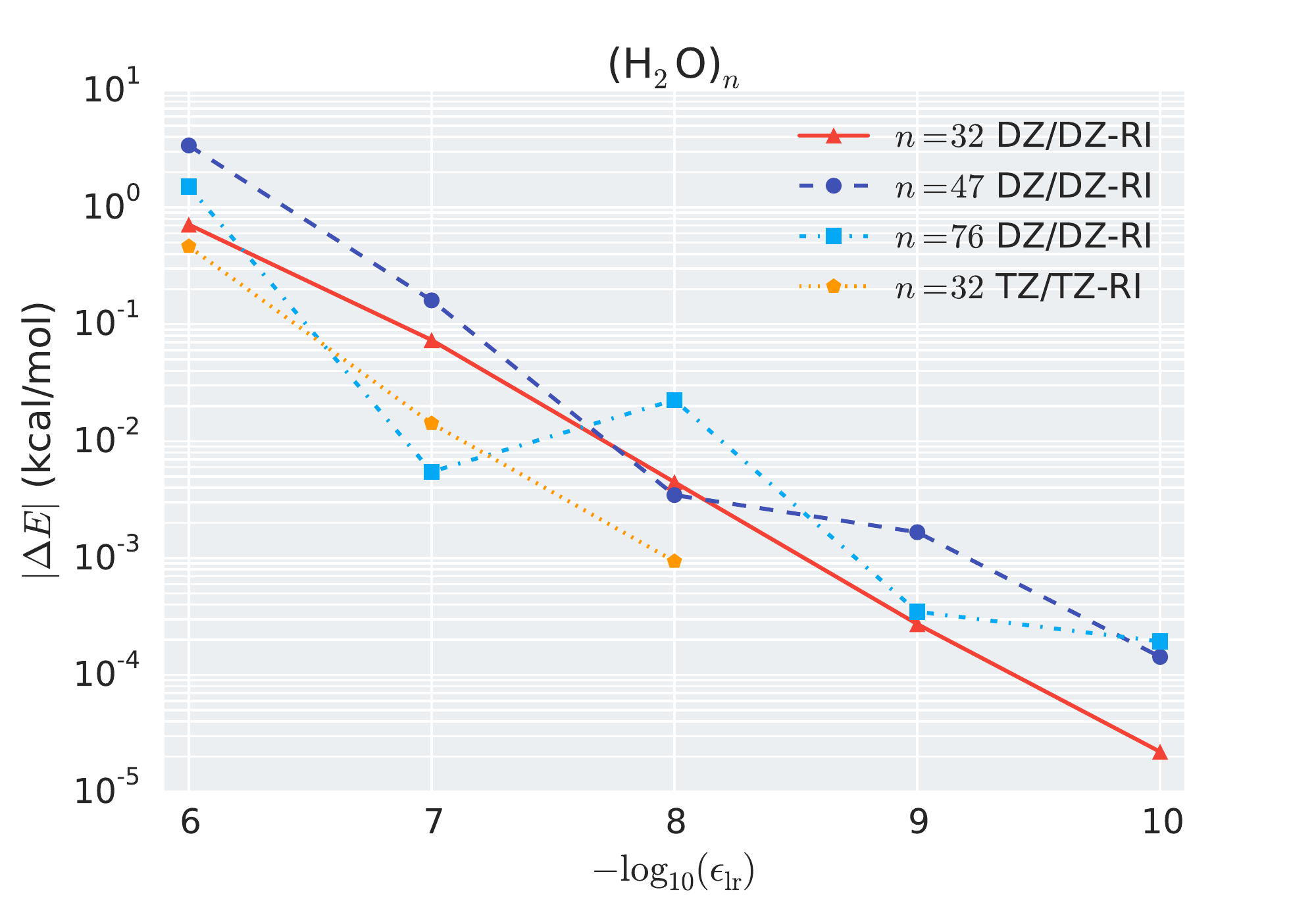}
        \caption{
            CLR approximation error in the Hartree-Fock energy of water clusters as
            a function of parameter \epsil{lr} (\epsil{sp}=0).
        }
        \label{fig:lr_thresh_energies}
    \end{figure}
    \begin{figure}[h!]
        \includegraphics[width=\columnwidth]{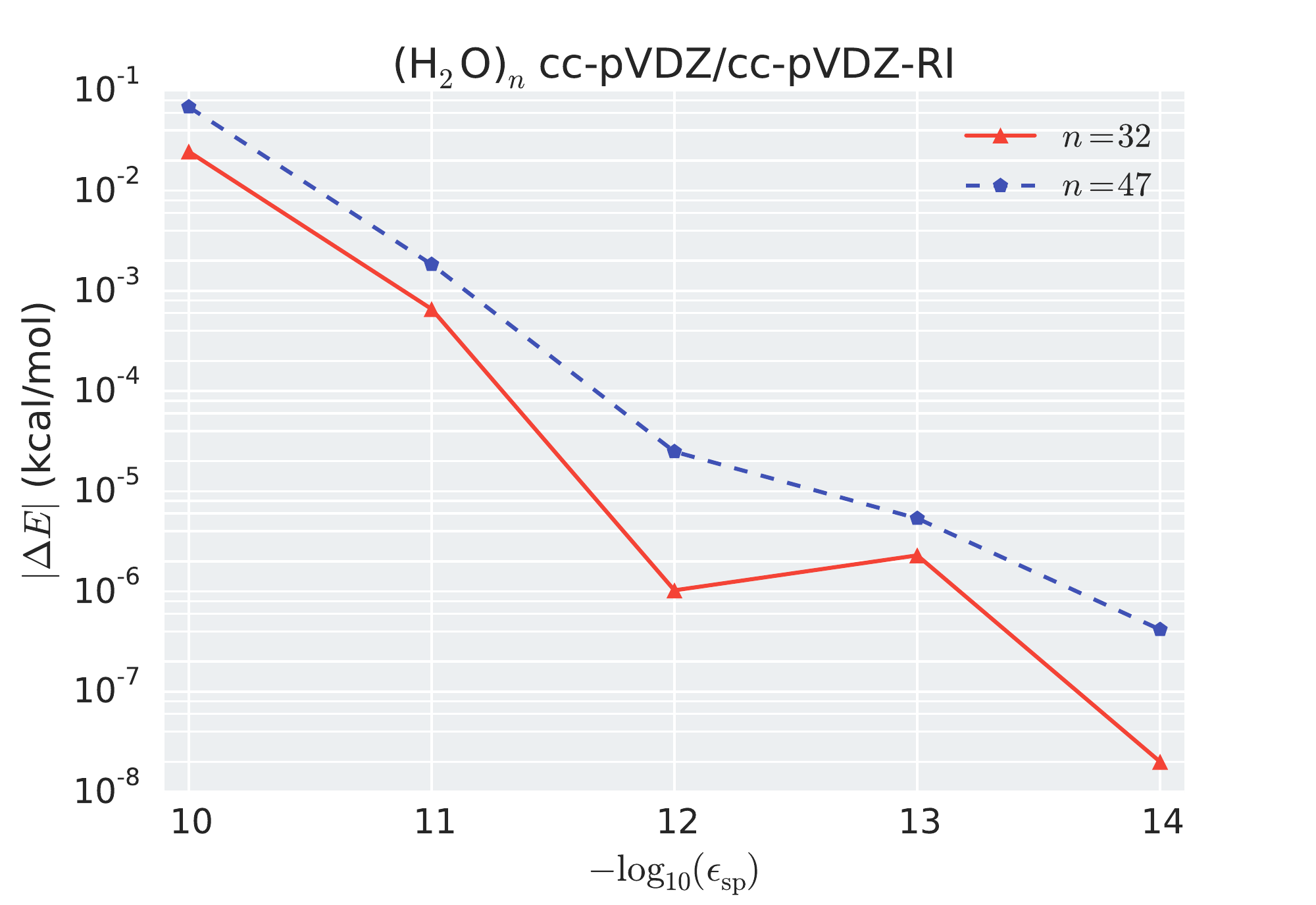}
        \caption{
            CLR approximation error in the cc-pVDZ Hartree-Fock energy of water clusters as
            a function of parameter \epsil{sp} (\epsil{lr}=0).
        }
        \label{fig:sp_thresh_energies}
    \end{figure}

\begin{figure}[h!]
        \includegraphics[width=\columnwidth]{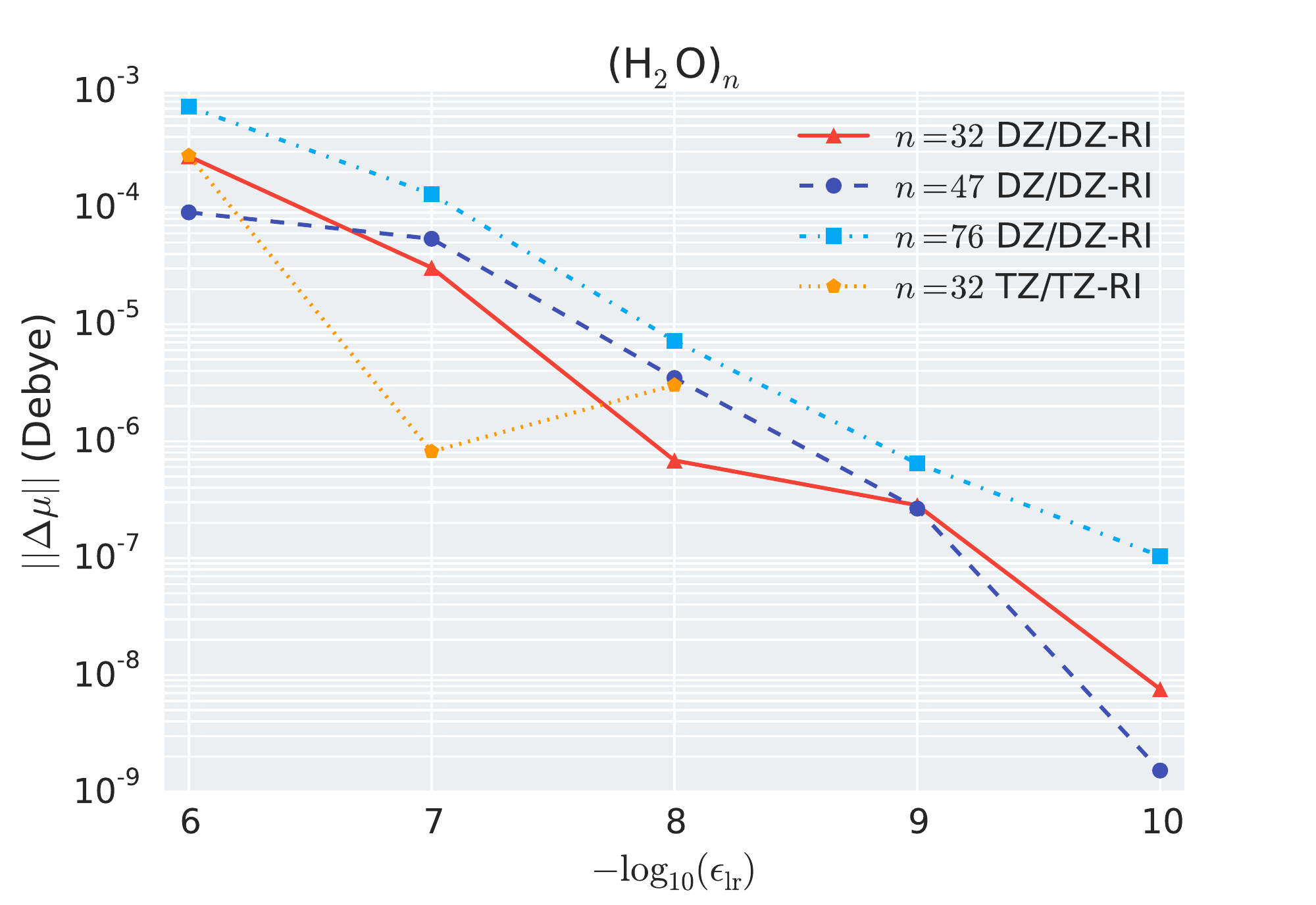}
        \caption{
            CLR approximation error in the Hartree-Fock dipole moment of water clusters as
            a function of parameter \epsil{lr} (\epsil{sp}=0).
        }
        \label{fig:lr_thresh_dipoles}
    \end{figure}

\begin{figure}[h!]
        \includegraphics[width=\columnwidth]{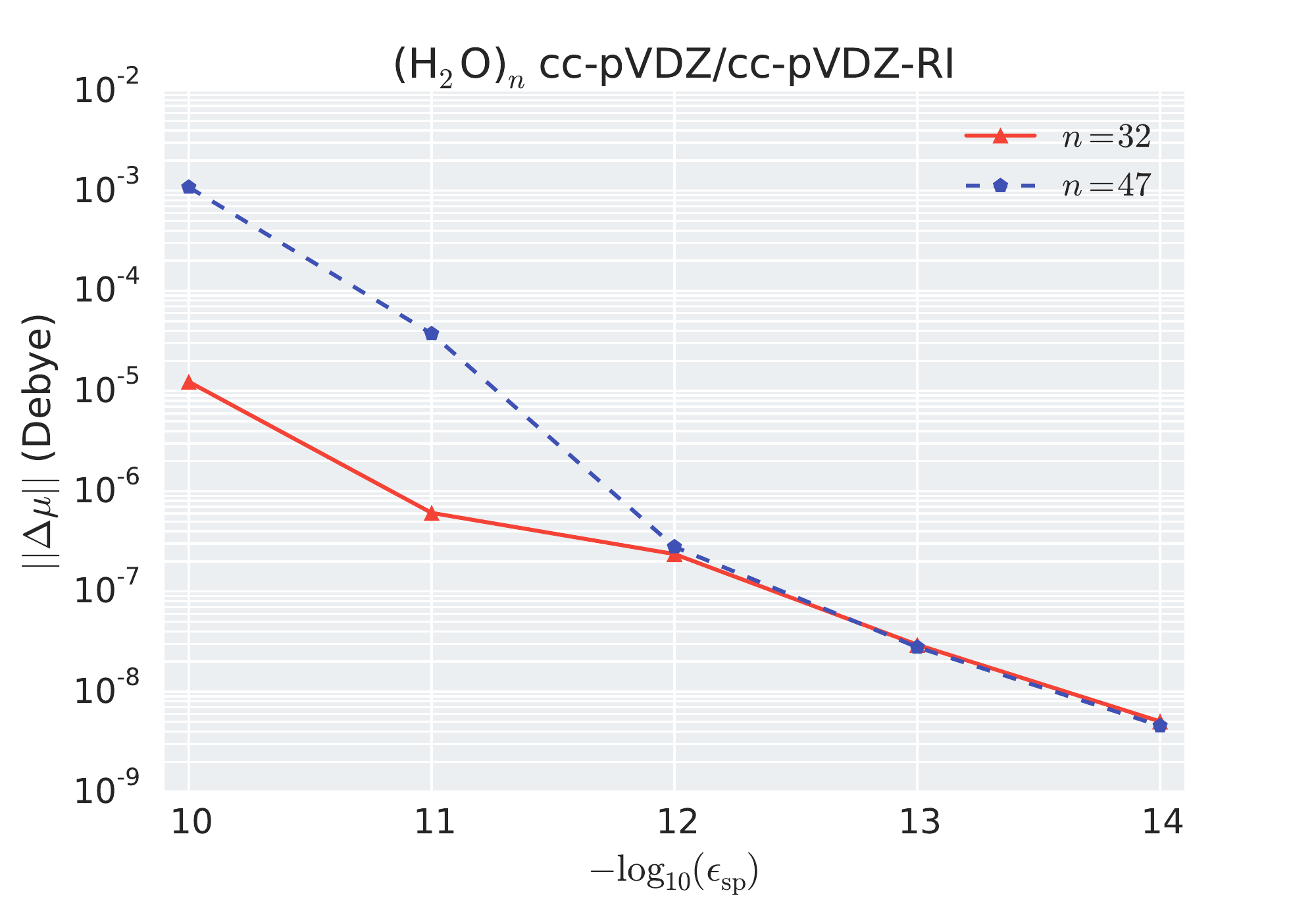}
        \caption{
            CLR approximation error in the cc-pVDZ Hartree-Fock dipole moment of water clusters as
            a function of parameter \epsil{sp} (\epsil{lr}=0).
        }
        \label{fig:sp_thresh_dipoles}
    \end{figure}

\begin{figure}[h!]
        \includegraphics[width=\columnwidth]{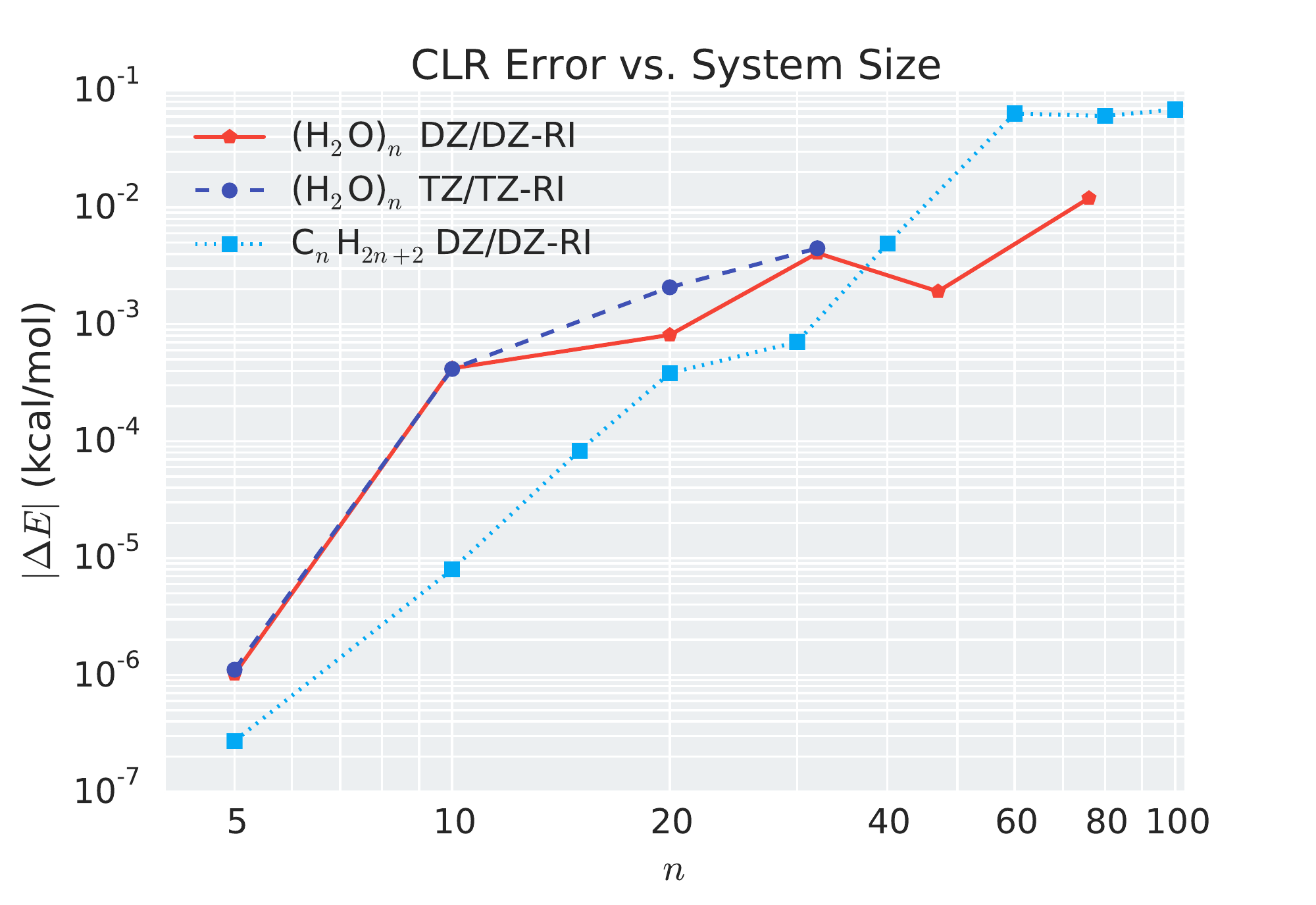}
        \caption{
            CLR approximation error in the Hartree-Fock energies of water clusters and n-alkanes as
            a function of system size ($\epsil{lr}=10^{-8}, \epsil{sp}=10^{-11}$).
        }
        \label{fig:error_en_sys_size}
    \end{figure}

\begin{figure}[h!]
        \includegraphics[width=\columnwidth]{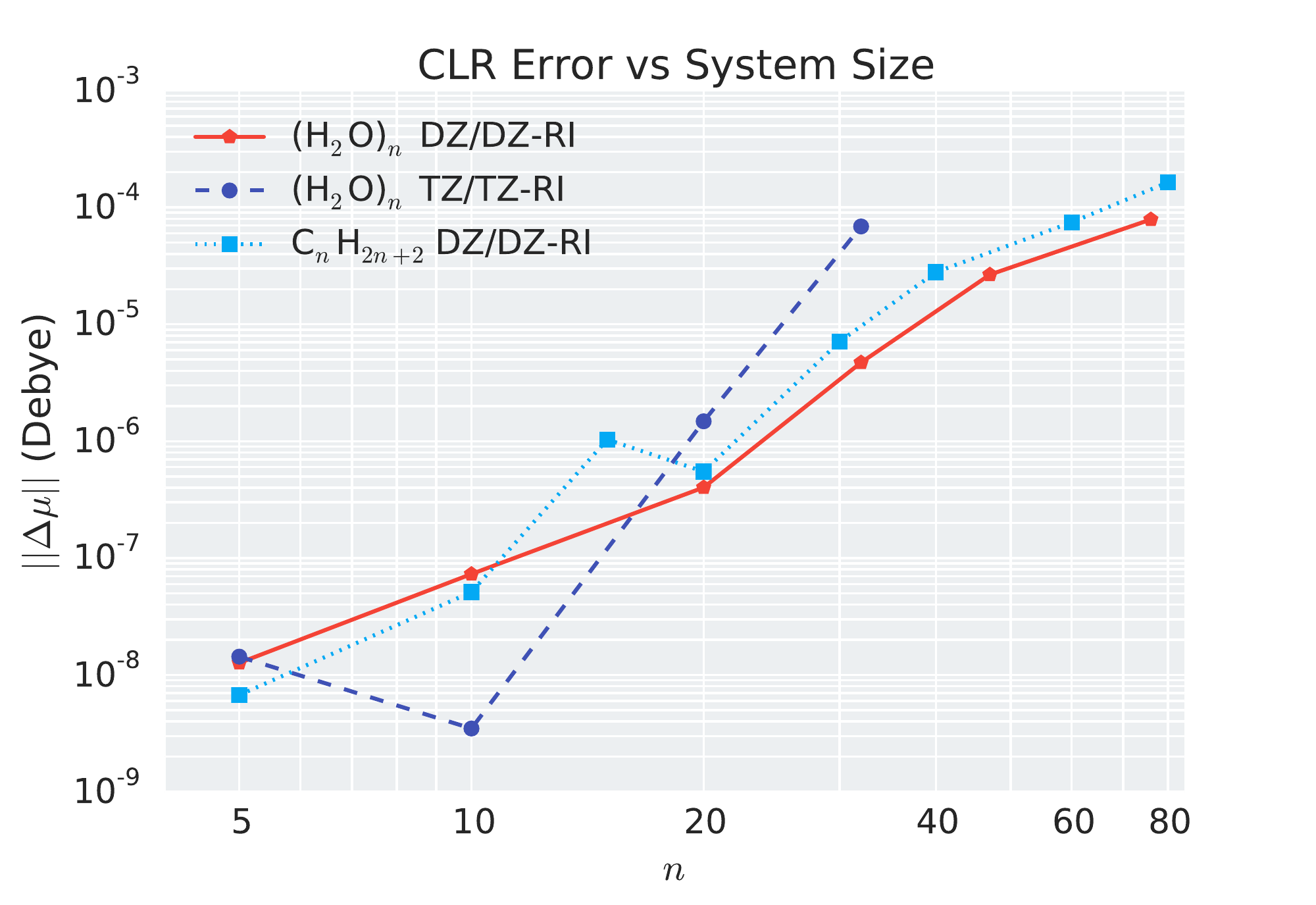}
        \caption{
            CLR approximation error in the Hartree-Fock dipole moments of water clusters and n-alkanes as
            a function of system size ($\epsil{lr}=10^{-8}, \epsil{sp}=10^{-11}$).
        }
        \label{fig:error_dp_sys_size}
    \end{figure}

    Figures \ref{fig:lr_thresh_energies}-\ref{fig:error_dp_sys_size} report the errors in absolute electronic energies and in electric
    dipole moments relative to the standard DF HF method:
    \begin{align}\label{eq:clr_errors}
        | \Delta E | &\equiv | E_{\mathrm{CLR}} - E_{\mathrm{ref}} |, \\
        | \Delta \boldsymbol{\mu} | & \equiv | \,||\boldsymbol{\mu}_{\mathrm{CLR}}|| - ||\boldsymbol{\mu}_{\mathrm{ref}}||\, | .
    \end{align}
	As the two CLR thresholds,
    \epsil{sp} and \epsil{lr} approach zero the error due to CLR should approach zero as well.
    Figures \ref{fig:lr_thresh_energies}-\ref{fig:sp_thresh_dipoles} demonstrate that this is indeed
    the case: the errors decay in proportion to the CLR truncation parameters.
    Although the errors do not decrease monotonically (e.g. CLR approximation
    does not guarantee variational character of the energy), the errors can be made arbitrarily small.
	The errors in energies are all well below the chemical accuracy target.
	It's clear that even the most crude threshold values, $\epsil{lr}=10^{-6}$ and $\epsil{sp}=10^{-10}$,
	will be sufficient for practical chemical applications of hybrid KS DFT. For the rest of the paper
	we use more stringent values of these parameters, $\epsil{lr}=10^{-8}$ and $\epsil{sp}=10^{-11}$,
	which should be appropriate for chemical-accuracy evaluation of correlation energies from the Hartree-Fock
	orbitals and potentials.
    

    Figures \ref{fig:error_en_sys_size} and \ref{fig:error_dp_sys_size}
    demonstrate the accuracy with regards to system size for the combined
    approximation.
    While total error increases with system size, the error per unit for
    example, while not constant, is controllable and for these calculations 
    is never above $2\times10^{-4}$ kcal/mol per water molecule or
    $1.1\times10^{-3}$ kcal/mol per CH$_{2}$ unit in n-alkanes.


\subsubsection{Assessment of Storage Reduction}
    \begin{figure}[h!]
        \includegraphics[width=\columnwidth]{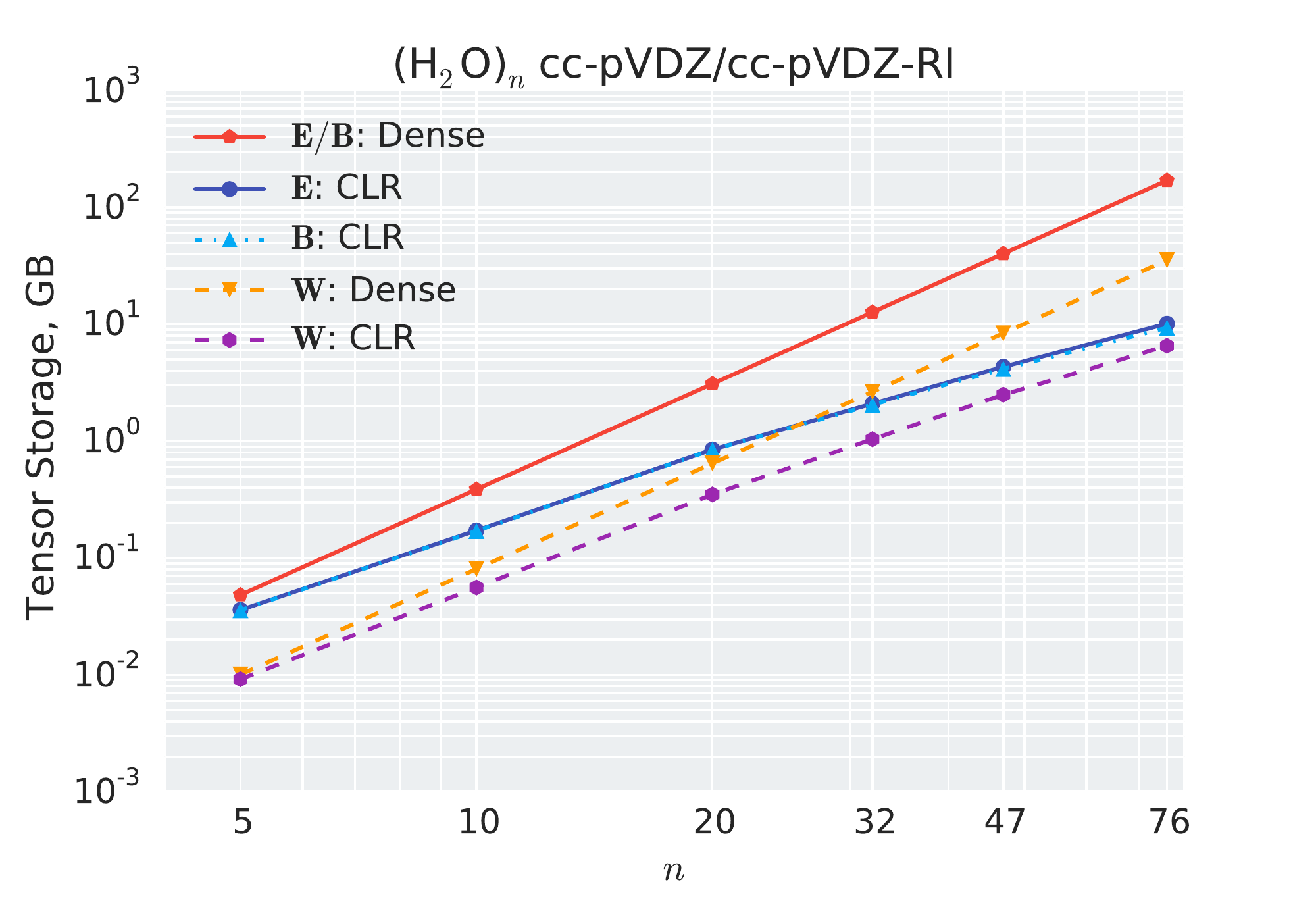}
        \caption{
            Size (GB) of key order-3 tensors evaluated in density-fitting-based Hartree-Fock method,
            with and without CLR approximation, as a function of the water cluster size.
            Basis=cc-pVDZ/cc-pVDZ-RI.
        }
        \label{fig:memory_scaling_water_dz}
    \end{figure}

    \begin{figure}[h!]
        \includegraphics[width=\columnwidth]{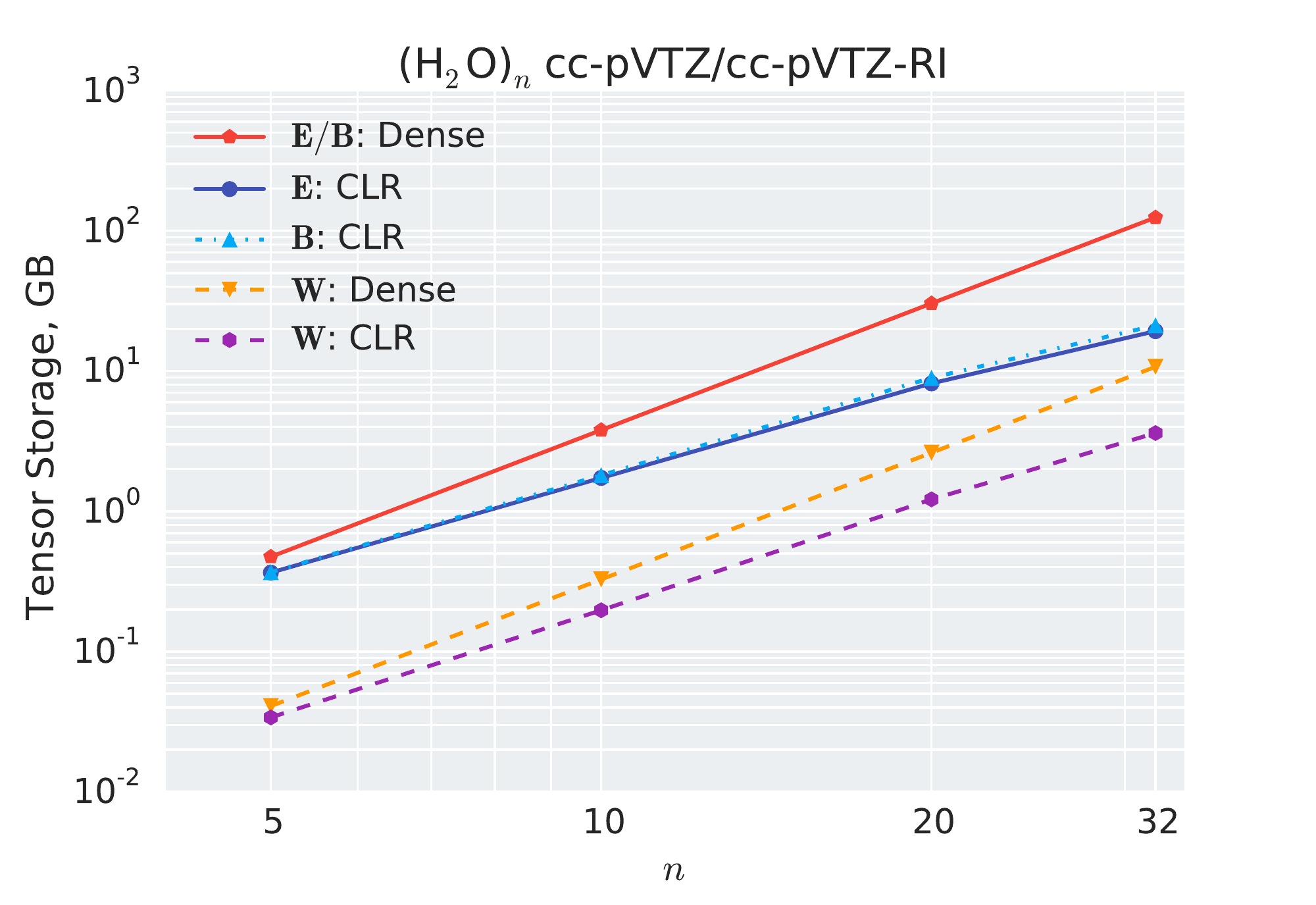}
        \caption{
            Size (GB) of key order-3 tensors evaluated in density-fitting-based Hartree-Fock method,
            with and without CLR approximation, as a function of the water cluster size.
            Basis=cc-pVTZ/cc-pVTZ-RI.
        }
        \label{fig:memory_scaling_water_tz}
    \end{figure}

    \begin{figure}[h!]
        \includegraphics[width=\columnwidth]{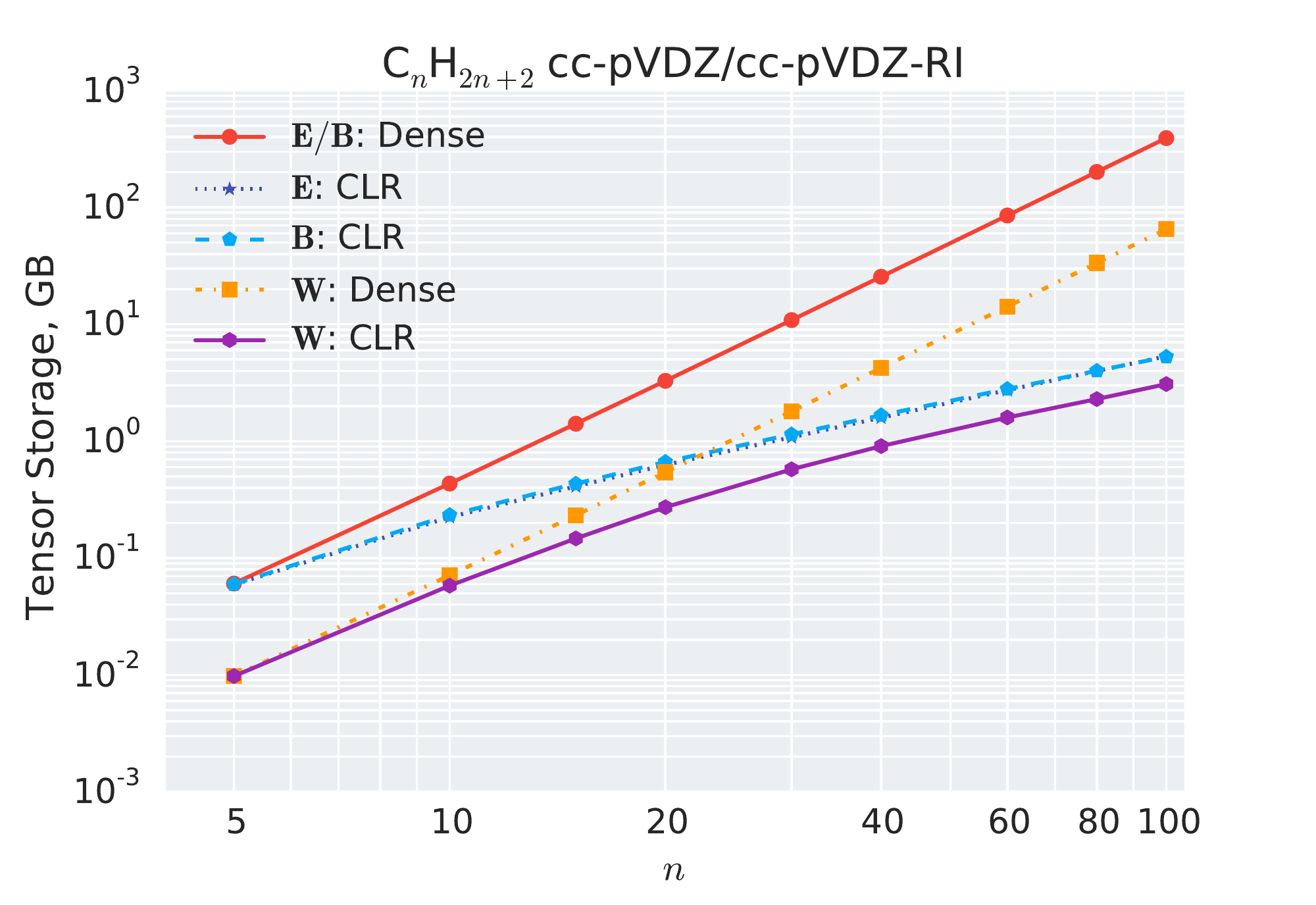}
        \caption{
            Size (GB) of key order-3 tensors evaluated in density-fitting-based Hartree-Fock method,
            with and without CLR approximation, as a function of the n-alkane size.
            Basis=cc-pVDZ/cc-pVDZ-RI.
        }
        \label{fig:memory_scaling_alkane_dz}
    \end{figure}

    Figures \ref{fig:memory_scaling_water_dz},
    \ref{fig:memory_scaling_water_tz}, and \ref{fig:memory_scaling_alkane_dz}
    show that the use of CLR approximation drastically reduces the data size of
    order-3 tensors that appear in the DF HF method (storage of the iteration-dependent $\mat{W}$ tensor
    was evaluated after 5 SCF iterations). For example, storage for
    tensors $\mat{E}$ and $\mat{B}$ in the double-$\zeta$ basis for
    (H$_{2}$O)$_{76}$ was reduced, using of CLR, by a factor 16.8 for $\mat{E}$
    and 18.2 for $\mat{B}$. For quasi-1-dimensional C$_{100}$H$_{202}$ these values were
    73.6, and 74.3, respectively.
    \footnote{
        Note that due to the lack of support for permutational symmetry in 
        {\sc TiledArray} (the feature is under current development) the storage
        size for both $\mat{E}$ and $\mat{B}$ is twice what it should be in
        practice, however this deficiency does not affect the storage
        comparison between CLR and standard DF.
    }
    
    The apparent storage complexity is also significantly reduced compared to \bigO{N^{3}} of the
    standard DF approach.
    We estimated the effective storage exponent
    by finite difference from the data for the two largest systems in each series.
    For n-alkanes the complexity estimates are \bigO{N^{1.30}} for $\mat{E}$, \bigO{N^{1.24}} for $\mat{B}$,
    and \bigO{N^{1.32}} for $\mat{W}$.
    Complexity reduction is observed for 3-dimensional systems as well, although they are not as drastic.
    For example, for water clusters the complexity estimates are
    \bigO{N^{1.77}} and \bigO{N^{1.82}} for the $\mat{E}$ tensor with the double- and triple-$\zeta$
    bases, respectively.

    Note that unlike the standard method, in which ratio of storage between
    tensors \mat{E}/\mat{B} and \mat{W} is constant, the CLR storage data
    reveals that the extent of data sparsity in tensors \mat{E} and \mat{B}
    grows faster than in \mat{W}. Using localized occupied orbitals rather than
    the Cholesky orbitals can potentially decrease the \mat{W} storage further,
    however detailed investigation of this issue is outside this preliminary
    investigation.

\subsubsection{Performance Assessment}
    \begin{figure}[h!]
        \includegraphics[width=\columnwidth]{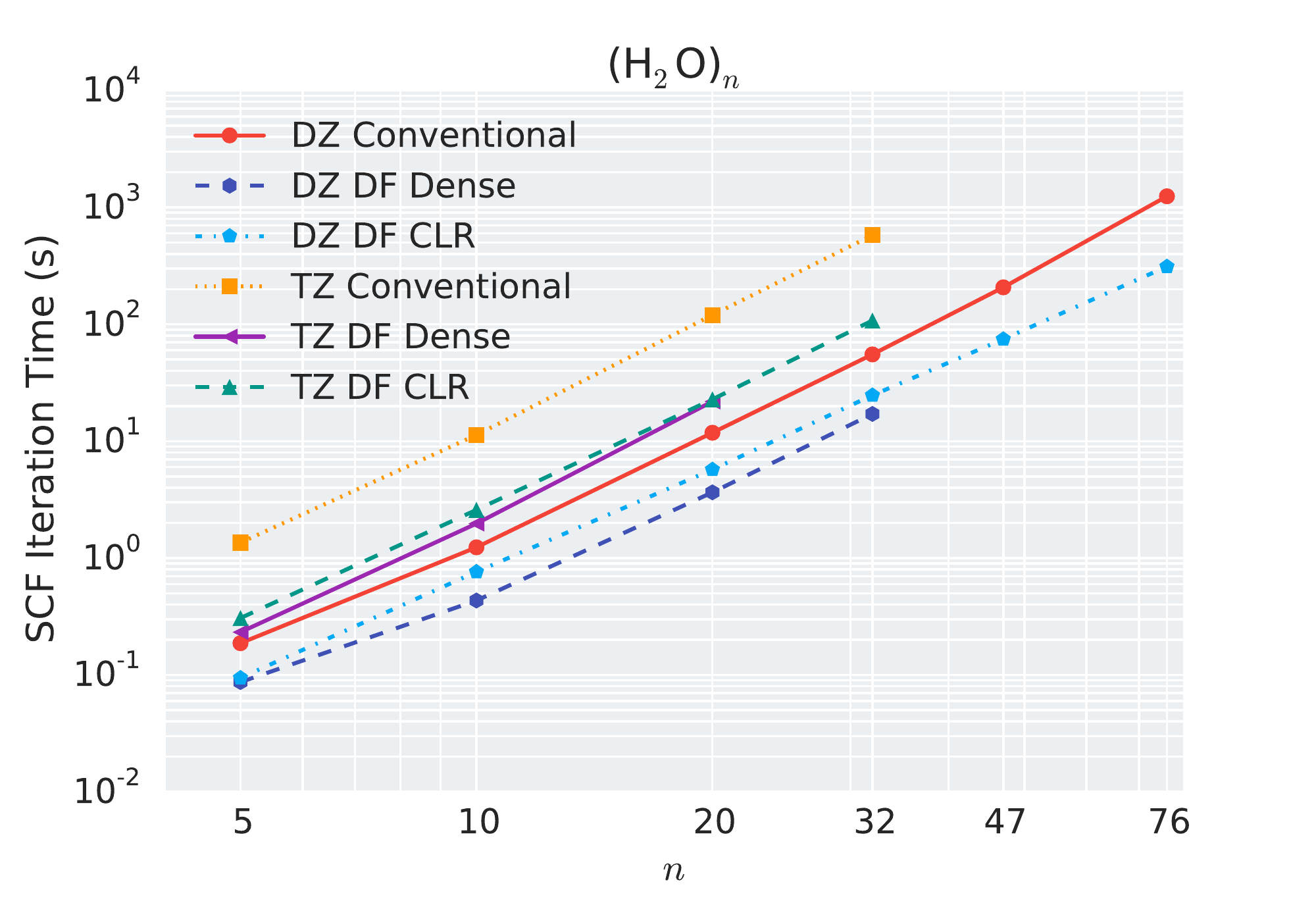}
        \caption{
            Average SCF iteration times for water clusters.
            \epsil{sp}$=1\times10^{-11}$ and \epsil{lr}$=1\times10^{-8}$.
        }
        \label{fig:water_scf_times}
    \end{figure}

    \begin{figure}[h!]
        \includegraphics[width=\columnwidth]{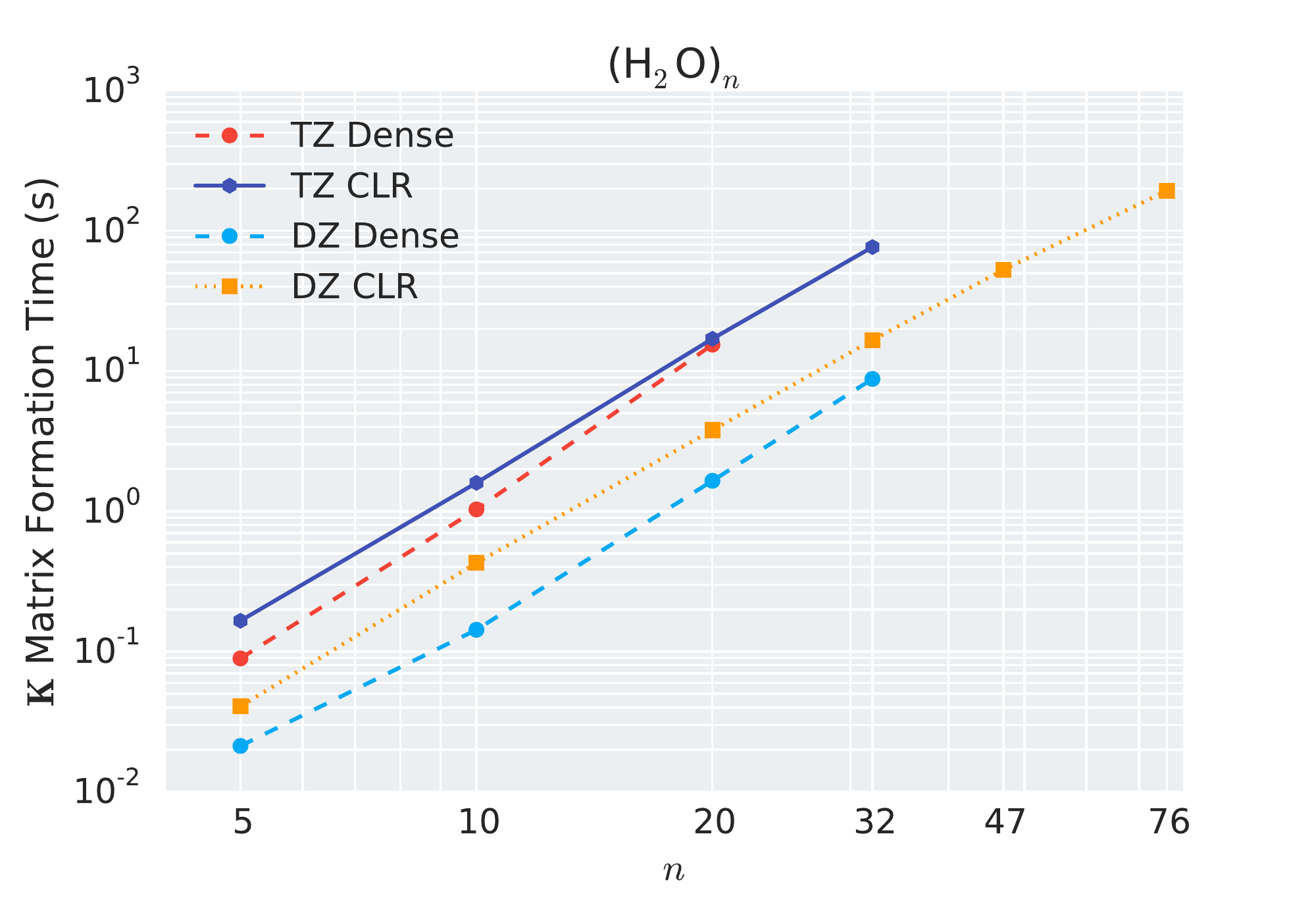}
        \caption{
            Average exchange build times for water clusters.
            \epsil{sp}$=1\times10^{-11}$ and \epsil{lr}$=1\times10^{-8}$.
        }
        \label{fig:water_k_times}
    \end{figure}
    
    Our implementation supports massive distributed memory
    parallelism in the Fock matrix build due to the use of {\sc TiledArray} framework.\cite{IA3-TA-paper}
    To simplify the performance assessment here we
    limited all tests to a single machine with two 8-core Intel Xeon
    E5-2650 v2 @ 2.60GHz processors and 64
    GB of DRAM. Parallel performance, as well as the analysis of performance of block sizes and other
    parameters will be explored elsewhere.

\begin{figure}[h!]
        \includegraphics[width=\columnwidth]{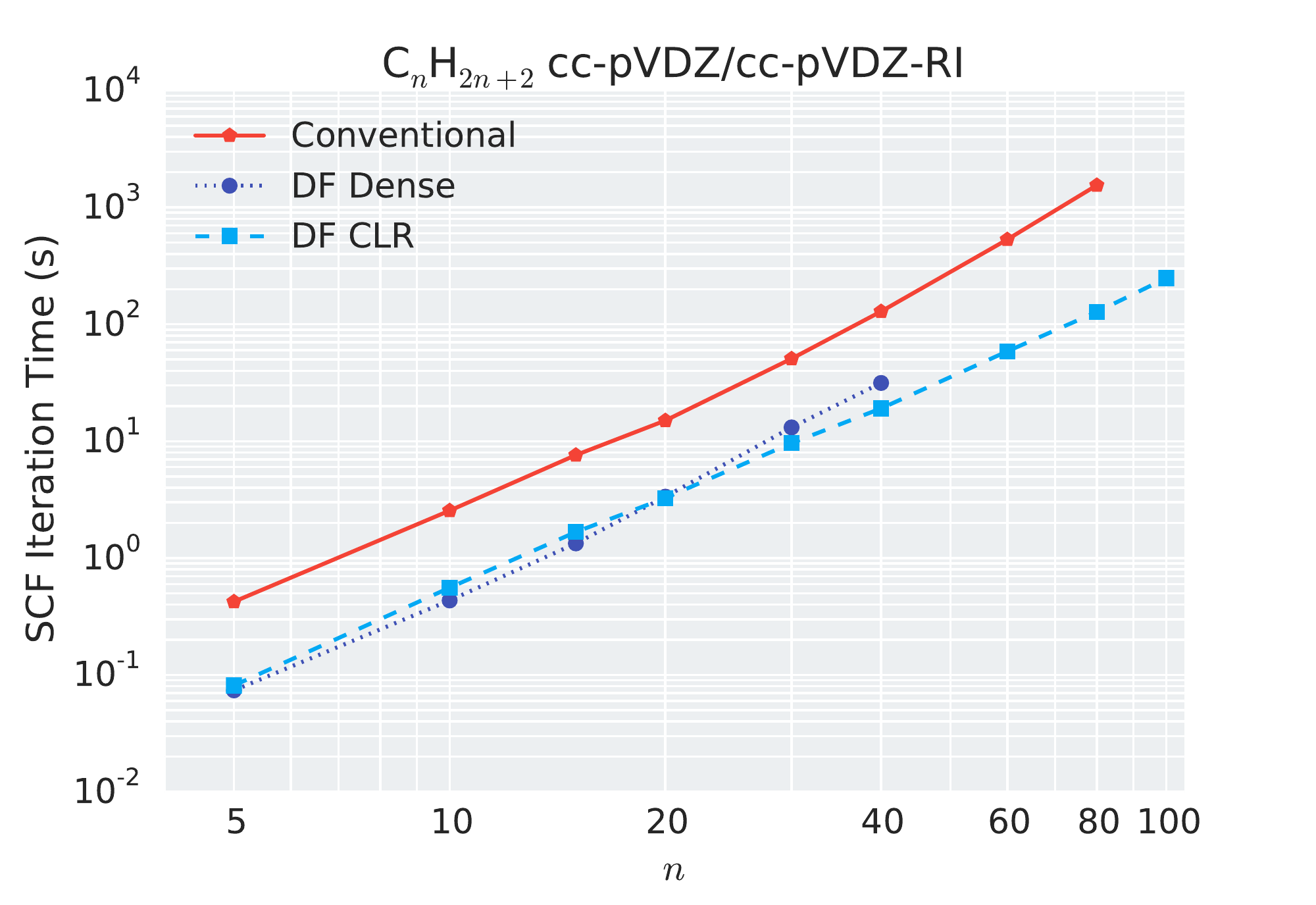}
        \caption{
            Average SCF iteration times for n-alkanes.
            \epsil{sp}$=1\times10^{-11}$ and \epsil{lr}$=1\times10^{-8}$.
        }
        \label{fig:alkane_scf_times}
    \end{figure}

Figures \ref{fig:water_scf_times} and \ref{fig:alkane_scf_times}
show average time per Hartree-Fock SCF iteration for water clusters and n-alkanes
obtained with the conventional, DF, and CLR DF approaches (for the former we used the parallel HF
test program of Libint library v. 2.1.0-beta). Both standard and CLR DF methods are significantly faster than
the conventional counterpart, even when double-$\zeta$ basis sets are used.
However, the standard DF approach quickly runs out of the 64 GB RAM:
memory allocation becomes problematic around (H$_2$O)$_{32}$ and C$_{40}$H$_{82}$ in a double-$\zeta$
basis and around (H$_2$O)$_{20}$ in a triple-$\zeta$ basis.
(conventional DF HF codes of course can avoid this bottleneck by using disk and/or reformulating the exchange build to
minimize the storage, but neither strategy reduces the storage complexity to below cubic). Due to the greatly reduced storage
requirement much larger computations are possible with the CLR DF method.

    \begin{figure}[h!]
        \includegraphics[width=\columnwidth]{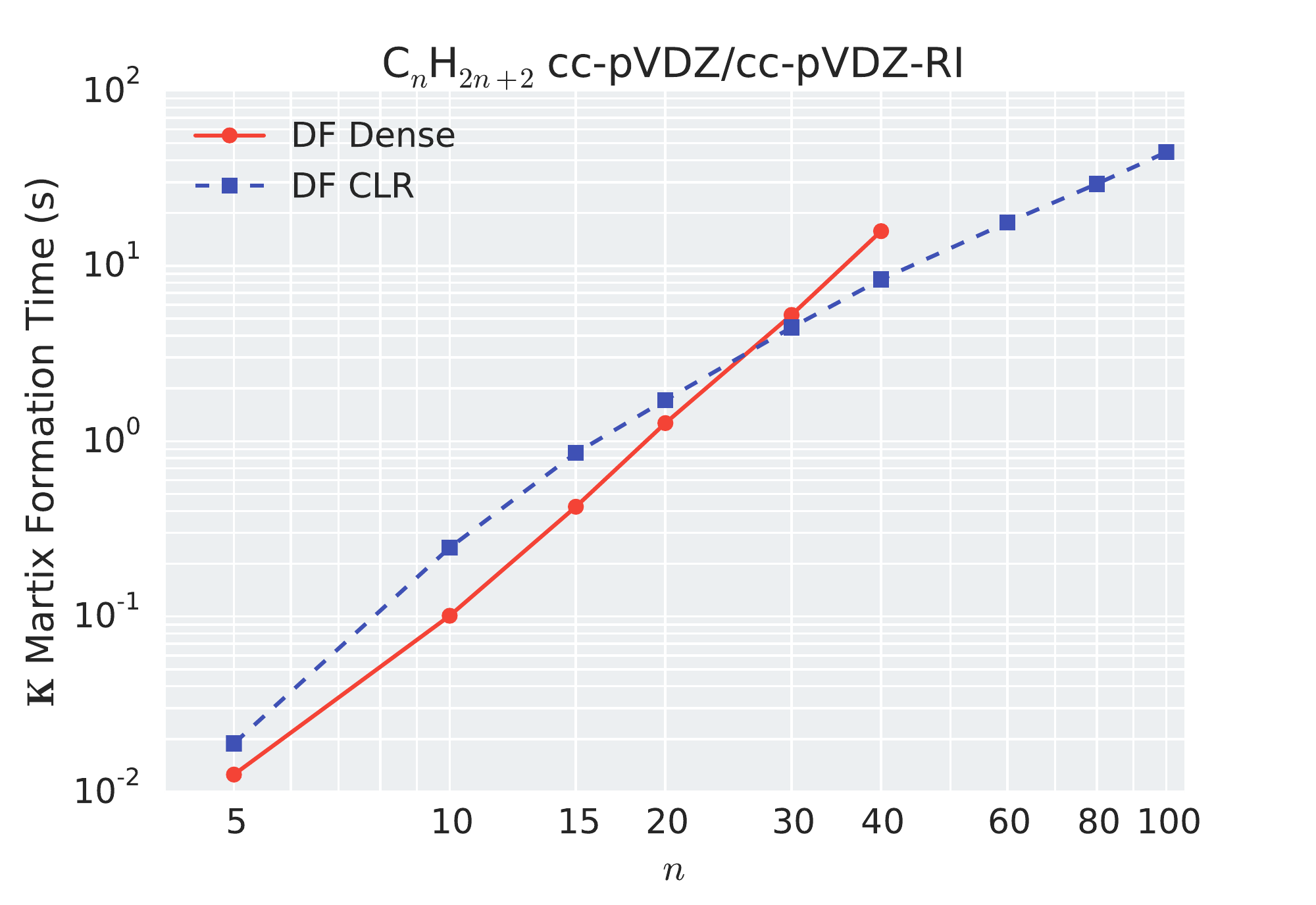}
        \caption{
            Average exchange build times for n-alkanes.
            \epsil{sp}$=1\times10^{-11}$ and \epsil{lr}$=1\times10^{-8}$.
        }
        \label{fig:alkane_k_times}
    \end{figure}

Most importantly,
for n-alkanes in a double-$\zeta$ basis
CLR DF HF becomes faster than its standard DF HF counterpart, due to lower {\em computational} complexity.
This is best demonstrated when considering timings for the most expensive step only, namely
the exchange matrix construction (this includes computation of \mat{W} and \mat{K}
matrices). Figure \ref{fig:alkane_k_times}
demonstrates that for double-$\zeta$ alkanes the CLR-based method becomes faster
than the standard counterpart between 20 and 30 carbon atoms.
For the triple-$\zeta$ water clusters the crossover occurs around 20 water molecules;
for the double-$\zeta$ basis the crossover does not occur, but due to the smaller slope
of the CLR line the cross-over is bound to occur for a larger system, with more than 50 water molecules.

We estimated the effective computational complexity of the exchange construction
using the finite difference from the two data points for the largest systems in each series.
For water clusters and n-alkanes
in a double-$\zeta$ basis the observed complexities
is \bigO{N^{2.7}} and \bigO{N^{1.86}}, respectively, and for water clusters
in a triple-$\zeta$ basis the complexity is \bigO{N^{3.2}}. All of these figures compare favorably
to the \bigO{N^{4}} complexity of the standard DF exchange algorithm.

\section{\label{sec:conclusions} Conclusions and Perspective}
    In this work we introduced the Clustered Low Rank (CLR)
    framework for block-sparse and block-low-rank tensor representation and computation.
    Use of the CLR format for the order-2 and order-3 tensors
    that appear in the context of density-fitting-based Hartree-Fock method
    significantly reduced the storage and computational complexities
    below their standard \bigO{N^{3}} and \bigO{N^{4}} figures. {\em Even for relatively small systems and realistic
    basis sets} CLR-based DF HF becomes more efficient
    than the standard DF approach while negligibly affecting molecular energies and properties.
    
    The entire computation framework that we described here depends on 2 parameters
    that control precision. Precision of the CLR representation depends on input parameter that controls
    the block rank truncation; as it approaches 0 the representation becomes exact.
    Another parameter controls screening of small contributions in
    arithmetic operations on CLR tensors; as it approaches 0 the 
    arithmetic becomes exact. There are no other ad-hoc heuristics,
    such as domains.

    This is an initial application of the CLR format, and many significant optimization opportunities remain to be explored.
	Nevertheless, the efficiency of CLR DF HF method immediately makes it useful on its own and as a building block for other
	reduced scaling methods. For example, the fast CLR-based DF methodology should also be immediately
	applicable in other contexts where density fitting is key,
    e.g. the reduced scaling electron correlation methods. The high efficiency of the CLR DF Fock build
    suggests that the analytic gradients for CLR-based hybrid
	KS DFT is a logical next step, to allow efficient on-the-fly dynamics.
	We should note that the CLR framework should be naturally beneficial for
	massive parallelism necessary for ab initio dynamics, since the CLR data compression should reduce the
	traffic through the memory hierarchy, whether between memory tiers in the next generation of ``accelerators''
	or through the network in a cluster.
	
    Last, but not least, it is exciting
    to imagine uses of the CLR framework for exploiting
    the data sparsity in other tensors that appear in electronic structure and related fields, 
    such as the wave function projections (e.g. cluster amplitudes), density matrices and Green's functions.
	Some work along these lines is already underway.
    
\section{Acknowledgements}
We would like to thank Fabijan Pavo\v{s}evi\'c for useful discussions and
Virginia Tech's Advanced Research Computing for providing time on BlueRidge cluster,
which was used for debugging and obtaining reference values for large dense calculations.
We acknowledge the support by the U.S. National Science Foundation (grants CHE-1362655 and ACI-1450262), and by Camille and Henry Dreyfus Foundation.

\bibliographystyle{aipnum4-1}
\bibliography{refs2,coderefs,arxiv_ref,refs-ev}

\end{document}